\documentclass[preprint,12pt]{elsarticle}
\usepackage{amsfonts, amssymb, amsmath, mathrsfs}
\usepackage[top=1in,bottom=1in,left=1.25in,right=1.25in]{geometry}
\journal{{\rm } }

\begin{document}
\newtheorem{lemma}{Lemma}[section]
\newtheorem{proposition}{Proposition}[section]
\newtheorem{theorem}{Theorem}[section]
\newtheorem{corollary}{Corollary}[section]
\newtheorem{example}{Example}[section]
\newtheorem{definition}{Definition}[section]
\newtheorem{remark}{Remark}[section]
\newtheorem{property}{Property}[section]
 \makeatletter
    \newcommand{\rmnum}[1]{\romannumeral #1}
    \newcommand{\Rmnum}[1]{\expandafter\@slowromancap\romannumeral #1@}
    \makeatother

\begin{frontmatter}
\title{Information algebra system of soft sets}

 \author[label1]{Guan Xuechong\corref{cor1}}
 \ead{guanxc@foxmail.com}
 \cortext[cor1]{Corresponding author at: College of Mathematical Science, Xuzhou Normal University,
China. Tel.: +86 18260715158.}
 \author[label2]{Li Yongming}
\ead{liyongm@snnu.edu.cn}
\address[label1]{College of Mathematical Science, Xuzhou Normal University, Xuzhou, 221116, China}
\address[label2]{College of Computer Science, Shaanxi Normal University, Xi'an, 710062, China}

\begin{abstract}

Information algebra is algebraic structure for local computation and inference.
Given an initial universe set and a parameter set, we show that a soft set system over them is an information algebra.
Moreover, in a soft set system,
the family of all soft sets with a finite parameter subset can form a compact information algebra.

\end{abstract}


\end{frontmatter}

\section{Introduction }\label{intro}
The information algebra system introduced by Shenoy \cite{VBS} was
inspired by the formulation of some basic axioms of local
computation and inference under uncertainty \cite{Computation}. It
gives a basic mathematical model for treating uncertainties in
information. Related studies \cite{Information,Lecture,Semiring}
showed that the framework of information algebras covers many
instances from constraint systems, Bayesian networks,
Dempster-Shafer belief functions to relational algebra, logic and
etc. Considering about the feasibility of information processing
with computer, Kohlas \cite{Information,Lecture} presented a special
information algebra with approximation structure called compact
information algebra recently.

On the other hand, Molodtsov \cite{first} initiated a novel concept,
which is called soft set, as a new mathematical tool for dealing
with uncertainties \cite{soft}. In fact, a soft set is a
parameterized family of subsets of a given universe set. The way of
parameterization in problem solving makes soft set theory convenient
and simple for application. Now it has been applied in several
directions, such as operations research \cite{operation,Jiang},
topology \cite{top1,top2,top3}, universal algebra \cite{group,
algebra,ideal, semirings}, especially decision-making
\cite{reduction, decision, Zhi,comment, adjustable}.

It is thus evident that information algebra theory and soft set
theory are both theoretical research tools for dealing with
non-deterministic phenomenon. To study relationships between them is
necessary. In this paper, we are concerned about the problem that
whether there exist the frameworks of information algebras or even
compact information algebras in soft sets. By choosing some
appropriate operators, we construct an information algebra system of
soft sets over an initial universe set and a parameter set.
Then we further prove that, in a soft set system, the family of soft
sets with a finite parameter subset can form a compact information
algebra.
These conclusions obtained in this paper demonstrate that soft set
systems are also the instances of information algebras.

\section{Preliminaries}
\label{sec:2}

In this section, first, we present some basic definitions about soft
sets and some notations in lattice theory.

Suppose that $(L, \leq)$ is a partially ordered set and $A\subseteq
L$. We write $\vee A$ and $\wedge A$ for the least upper bound and
the greatest lower bound of $A$ in $L$ respectively if they exist.

Let $L$ be a partially ordered set. If $a\vee b$ and $a\wedge b$
exist for all $a,b\in L$, then we call $L$ a lattice. If $\vee A$
exists for every subset $A\subseteq L$, we call $L$ a complete
lattice. Clearly, a partially ordered set $L$ is a complete lattice
if, and only if, $L$ has the bottom element and $\vee A$ exists for
all nonempty subset $A\subseteq L$.

A set $A\subseteq L$ is said to be directed, if for all $a,b\in A$,
there is a $c\in A$ such that $a,b\leq c$. For $a,b\in L$, we call
$a$ way-below $b$, in symbols $a\ll b$, if and only if for all
directed subsets $X \subseteq L$, if $\vee X$exists and $b\leq \vee
X$, then there exists an $x\in X$ such that $a\leq x$.

Let $U$ be an initial universe set and $E$ be a set of parameters,
which usually are initial attributes, characteristics, or properties
of objects in the initial universe set. ${\cal P}(U)$ denotes the
power set of $U$.

\begin{definition}\label{definition:1} {\rm (\cite{first})}
{\rm A pair $(F,A)$ is called a soft set over $U$, where $F$ is a
mapping given by $F : A\rightarrow {\cal P}(U)$. }
\end{definition}

Therefore a soft set is a tuple which associates with a set of
parameters and a mapping from the parameter set into the power set
of an universe set. In other words, a soft set over $U$ is a
parameterized family of subsets of the universe $U$. For
$\varepsilon\in A$, $F(\varepsilon)$ may be considered as the set of
$\varepsilon$-approximate elements of the soft set
$(F,A)$\cite{first}.

\begin{definition}\label{definition:4}{\rm (\cite{soft})}
{\rm A soft set $(F, A)$ over $U$ is said to be a null soft set, if
for all $e\in A, F(e)=\emptyset$. We write it by $(\emptyset, A)$.

A soft set $(F, A)$ over $U$ is said to be an absolute soft set
denoted by $\tilde{A}$, if for all $e\in A, F(e)= U$.}
\end{definition}

\begin{definition}\label{definition:6}{\rm (\cite{operation})}
{\rm The extended intersection of two soft sets $(F, A)$ and $(G,
B)$ over a common universe $U$ is the soft set $(H, C)$, where $C =
A \cup B$, and $\forall e \in C$,
$$H(e)=\left\{
          \begin {array}{ll}
          F(e),&{\mbox{if}} \ \ e\in A-B;\\
          G(e),&{\mbox{if}} \ \ e\in B-A;\\
          F(e)\cap G(e),&{\mbox{if}} \ \ e\in A\cap B.
         \end{array}
        \right.$$
We write $(F,A)\sqcap_{\varepsilon} (G,B)=(H,C)$.}
\end{definition}

In this paper we adopt the concept of information algebra given by
Kohlas from \cite{Information}. For a full introduction and these
abundant examples of information algebras, please refer to
\cite{Information,Lecture,Semiring}.

\begin{definition}\label{definition:14}{\rm (\cite{Information})}
{\rm Let $(D, \leq)$ be a lattice. Suppose there are three
operations defined in the tuple $(\Phi, D)$:

1.Labeling $d$: $\Phi \rightarrow D; \phi\mapsto d(\phi)$, where
$d(\phi)$ is called the domain of $\phi$. For an $s\in D$, let
$\Phi_s$ denote the set of all valuations with domain $s$.

2.Combination $\otimes$: $\Phi\times \Phi \rightarrow \Phi;
(\phi,\psi)\mapsto \phi\otimes\psi$,

3.Marginalization $\downarrow$: $\Phi\times D \rightarrow \Phi;
(\phi,x)\mapsto \phi^{\downarrow x}$, for $x\leq d(\phi)$.

If the system $(\Phi,D)$ satisfies the following axioms, it is
called an information algebra:

1.Semigroup: $\Phi$ is associative and commutative under
combination. For all $s\in D$, there is a neutral element $e_s$ with
$d(e_s)=s$ such that for all $\phi\in \Phi$ with $d(\phi)=s,
e_s\otimes \phi=\phi$.

2. Labeling: For $\phi,\psi \in \Phi$, $d(\phi\otimes\psi) =
d(\phi)\vee d(\psi)$.

3. Marginalization: For $\phi\in \Phi, x\in D, x\leq d(\phi),
d(\phi^{\downarrow x})=x$.

4. Transitivity: For $\phi\in \Phi$ and $x\leq y\leq d(\phi),
(\phi^{\downarrow y})^{\downarrow x}=\phi^{\downarrow x}$

5. Combination: For $\phi,\psi \in \Phi$ with $d(\phi)=x, d(\psi)=y,
(\phi\otimes\psi)^{\downarrow x}=\phi\otimes\psi^{\downarrow x\wedge
y}$.

6. Stability: For $x,y\in D, x\leq y$, $e_y^{\downarrow x}=e_x$.

7. Idempotency: For $\phi\in \Phi$ and $x\in D, x\leq d(\phi)$,
$\phi\otimes \phi^{\downarrow x}=\phi$. }
\end{definition}

The items putting forward in the definition of information algebra
can be seen as the axiomatic presentations of some basic principles
in local computation and inference. Studies have shown this
algebraic structure covers many instances from belief functions,
constraint systems, relational databases, and possibility theory to
relational algebra and logic(\cite{Semiring}). For example, each
lattice $L$ is a simply information algebra on a domain set $L$
itself. The operations are defined as follows:

1. Labeling $d$: For $x\in L$, $d(x)=x$.

2. Combination $\otimes$: $x\otimes y= x\vee y$.

3. Projection $\downarrow$: If $x\leq y$, $x^{\downarrow y}=x\wedge y$.\\

For an information algebra $(\Phi,D)$, we introduce a order relation
as follows:
\begin{center}
$\psi\leq \phi$, if $\psi\otimes \phi=\phi$.
\end{center}
This order relation induced by the operation combination is a
partial order on the set $\Phi$, if $(\Phi,D)$ is an information
algebra.

\section{Information algebra of soft sets}

In this section, with these operations of soft sets defined above,
we will construct an information algebra of soft sets. Let $U$ be an
initial universe set and $E$ be a set of parameters. ${\cal
S}_{U,E}$(or simply ${\cal S}$ when this doesn't lead to confusions)
denotes the set of all soft sets $(F,A)$ over $U$, where $A\subseteq
E$, that is,
\begin{center}
${\cal S}$=$\{(F,A): (F,A) \mbox{\ is\ a\ soft\ set\ over\ $U$,\
where $A\subseteq E$}\}.$
\end{center}

Three operations are defined as follows:

1. Labeling $d$: For a soft set $(F, A)$, we define $d((F, A))=A$.

2. Projection $\downarrow$: If $B\subseteq A$, we define $(F,
A)^{\downarrow B}$ to be a soft set $(G,B)$ such that for all $b\in
B$, $G(b)=F(b)$.

3. Combination $\otimes$: For any two soft sets $(F, A), (G, B)\in
{\cal S}$, we define $$(F, A)\otimes (G, B)= (F,
A)\sqcap_\varepsilon (G, B).$$

We call a quintuple $({\cal S}, {\cal P}(E), d, \sqcap_\varepsilon,
\downarrow)$(abbreviated as $({\cal S}, {\cal P}(E))$ a soft set
system over $U$ and $E$. Now we show this system is an information
algebra.

\begin{theorem}\label{theorem:8}
{\rm The soft set system $({\cal S}, {\cal P}(E))$ over $U$ and $E$
is an information algebra. }
\end{theorem}
\noindent {\bf Proof.} Obviously, ${\cal P}(E)$ is a lattice
composed by the domains of soft sets in ${\cal S}$.

1. Semigroup: Clearly ${\cal S}$ is commutative with respect to the
operation $\sqcap_\varepsilon$. For $A\subseteq E$, the absolute
soft set $\tilde{A}$ is the neutral element such that
$(F,A)\sqcap_\varepsilon \tilde{A}=(F,A)$ for all soft set $(F,A)$
with domain $A$.

Following we show the associative law holds in the set ${\cal S}$.
Let $(F,A), (G,B), (H,C)\in {\cal S}$. We write

\begin{center}
$(F,A)\sqcap_\varepsilon(G,B)=(Q_1, A\cup B)$,\\
$(G,B)\sqcap_\varepsilon(H,C)=(Q_2, B\cup C)$,\\
$[(F,A)\sqcap_\varepsilon(G,B)]\sqcap_\varepsilon(H,C)=(Q_3, A\cup B\cup C)$,\\
$(F,A)\sqcap_\varepsilon[(G,B)\sqcap_\varepsilon(H,C)]=(Q_4, A\cup
B\cup C)$.
\end{center}
We need to show that $Q_3=Q_4$. For any an $e\in A\cup B\cup C$, it
can be divided into seven conditions as follows: $e\in (A-B)-C, e\in
(B-A)-C, e\in (A\cap B)-C, e\in C-(A\cup B), e\in (A-B)\cap C, e\in
(B-A)\cap C$ and $e\in A\cap B\cap C$. Here we take the condition of
$e\in (A\cap B)-C$ as an example to illuminate the proof. Assume
that $e\in (A\cap B)-C$. Since $(A\cap B)-C=A\cap (B-C)$, we have
$e\in A\cap (B-C)$. Then $$Q_3(e)=Q_1(e)=F(e)\cap G(e),$$ and
$$Q_4(e)=F(e)\cap Q_2(e)=F(e)\cap G(e).$$ So $Q_3(e)=Q_4(e)$. The
other conditions are also easy to show. Therefore, the associative
law holds.

2. According to these related definitions, the proof of the axioms
of labeling, marginalization, transitivity and idempotency are
directly.

3. Stability: For $A\in {\cal P}(E)$, the neutral element with
domain $A$ is the absolute soft set $\tilde{A}$. Furthermore, if
$B\subseteq A$, we have $\tilde{A} ^{\downarrow B}=\tilde{B}$. Thus
the stability is true.

4. Combination: For $(F,A), (G,B)\in {\cal S}$, we need to show
$$((F,A) \sqcap_\varepsilon (G,B))^{ \downarrow S} = (F,A)\sqcap_\varepsilon (G,B)^{\downarrow {S \cap B}},$$
if $A\subseteq S\subseteq A\cup B$.

In fact, let
$$(F,A) \sqcap_\varepsilon (G,B)=(H, A\cup B),$$
$$(F,A)\sqcap_\varepsilon (G,B)^{\downarrow {S \cap B}}=(H^{'}, S).$$
For all $e\in S$, we have
$$H(e)=H^{'}(e)=\left\{
          \begin {array}{ll}
          F(e),&{\mbox{if}} \ \ e\in S\cap(A-B);\\
          G(e),&{\mbox{if}} \ \ e\in S\cap(B-A);\\
          F(e)\cap G(e),&{\mbox{if}} \ \ e\in A\cap B.
          \end{array}
          \right.
          $$
Then $((F,A) \sqcap_\varepsilon (G,B))^{ \downarrow S} =
(F,A)\sqcap_\varepsilon (G,B)^{\downarrow {S \cap B}}$.

Hence $({\cal S}, {\cal P}(E), d, \sqcap_\varepsilon, \downarrow)$
is an information algebra. \qed

\section{Compact information algebra of soft sets}

In general, only ``finite" information can be treated in computers.
Therefore, a structure called compact information algebra has been
proposed by Kohlas. Its main character is that each information can
be approximated by these ``finite" information with a same domain.

\begin{definition}\label{definition:15}{\rm (\cite{Lecture})}
{\rm A system $(\Phi, \Phi_f, D)$, where $(\Phi,D)$ is an
information algebra, the lattice $D$ has a top element,
$$\Phi_f=\mathop\bigcup\limits_{x\in D} \Phi_{f,x}$$
where the sets $\Phi_{f,x}\subseteq \Phi_x$ are closed under
combination, contain the neutral element $e_x\in \Phi_{f,x}$, and
satisfy the following axioms of convergence and density with respect
to the ordering relation $\leq$ induced by the operation
combination, is called a compact information algebra.

1. Convergency: If $X\subseteq \Phi_{f,x}$ is directed, then the
supremum $\vee X$ over $\Phi$ exists and $\vee X\in \Phi_x$.

2. Density: For all $\phi\in \Phi_x$,
$$\phi=\bigvee \{\psi\in \Phi_{f,x}: \psi\leq \phi\}.$$

3. Compactness: If $X\subseteq \Phi_{f,x}$ is a directed set, and
$\phi\in \Phi_{f,x}$ such that $\phi\leq \vee X$ then there exists a
$\psi\in X$ such that $\phi\leq \psi$. }
\end{definition}

\begin{lemma}\label{lemma:3}{\rm (\cite{Lecture})}
{\rm If $(\Phi, \Phi_f, D)$ is a compact information algebra, then
$\phi\in \Phi_{f,x}$ if, and only if $\phi\ll \phi$ in set $\Phi_x$.
}
\end{lemma}

For convenience, we give an equivalent form for the order relation
$\leq$ induced by the operation combination of soft sets.

\begin{proposition}\label{proposition:1}
{\rm Let the order relation $\leq$ be induced by the operation
combination in the system $({\cal S}, {\cal P}(E))$. For two soft
sets $(F,A)$ and $(G,B)$ over a common universe $U$, then $(F,A)\leq
(G,B)$ if and only if,

(i) $A \subseteq B$, and

(ii) $\forall \varepsilon\in A$, $G(\varepsilon) \subseteq
F(\varepsilon)$. }
\end{proposition}
\noindent {\bf Proof.} We write $(F,A)\sqcap_{\varepsilon} (G,B)=(H,
A\cup B)$.

If $(F,A)\leq (G,B)$, then $(H, A\cup B)=(G,B)$. So $A\cup B=B$,
that is, $A \subseteq B$. For all $\varepsilon\in A$, by the
definition of the operation $\sqcap_{\varepsilon}$, we have
$H(\varepsilon)=F(\varepsilon)\cap G(\varepsilon)=G(\varepsilon)$.
Then $G(\varepsilon) \subseteq F(\varepsilon)$ for all
$\varepsilon\in A$.

The reverse is also obvious. \qed

\begin{proposition}\label{proposition:3}
{\rm Let $\{(F_i, A_i): i\in I\}$ be soft sets over a same universe
$U$. Then
$$\mathop\bigvee\limits_{i\in I}(F_i, A_i) = (H, \mathop\bigcup\limits_{i\in I} A_i),$$
where $H: \mathop\bigcup\limits_{i\in I} A_i\rightarrow {\cal P}(U)$
is defined as follows:

$\forall e\in \mathop\bigcup\limits_{i\in I} A_i$, let
$J^{(e)}=\{i\in I: e\in A_i\}$, $H(e)=\mathop\bigcap\limits_{i\in
J^{(e)}} F_i(e)$. }
\end{proposition}
\noindent {\bf Proof.} Clearly $(H,\mathop\bigcup\limits_{i\in I}
A_i)$ is an upper bound of $\{(F_i, A_i): i\in I\}$. Suppose that
$(G,B)$ is another upper bound of $\{(F_i, A_i): i\in I\}$. Thus
$\mathop\bigcup\limits_{i\in I} A_i\subseteq B$. $\forall e\in
\mathop\bigcup\limits_{i\in I} A_i, i\in J^{(e)}$, we have
$G(e)\subseteq F_i(e)$. Then $$G(e)\subseteq
\mathop\bigcap\limits_{i\in J^{(e)}} F_i(e)=H(e).$$ This proves that
$(H,\mathop\bigcup\limits_{i\in I} A_i)\leq (G,B)$. Thus
$\mathop\bigvee\limits_{i\in I}(F_i, A_i) = (H,
\mathop\bigcup\limits_{i\in I} A_i)$. \qed

\begin{proposition}\label{proposition:2}
{\rm  $({\cal S}_A,\leq)$ is a complete lattice. The top element is
$(\emptyset, A)$, and the bottom element is $\tilde{A}$. Here ${\cal
S}_A$ is the set of all soft sets with domain $A$ in the system
$({\cal S}, {\cal P}(E))$. }
\end{proposition}
\noindent {\bf Proof.} For all nonempty subset $\{(F_i,A): i\in
I\}\subseteq {\cal S}_A$, by the conclusion of Proposition
\ref{proposition:3}, we have
$$\mathop\bigvee\limits_{i\in I}(F_i, A) = (H, A)\in {\cal S}_A,$$
where $H: A\rightarrow {\cal P}(U)$ is defined as
$H(e)=\mathop\bigcap\limits_{i\in I} F_i(e)$ for all $e\in A$.
Moreover, $\tilde{A}$ is the bottom element in the set ${\cal S}_A$.
Thus ${\cal S}_A$ is a complete lattice.

\begin{lemma}\label{lemma:2}
{\rm  Let $(F,A)$ be a soft set over a universe $U$ and $A$ be a
finite subset of $E$. Then $(F,A)\ll (F,A)$ in ${\cal S}_{A}$ if,
and only if $U-F(e)$ is a finite subset of $U$ for all $e\in A$. }
\end{lemma}
\noindent {\bf Proof.} (1) ``if": Let $\{(G_i,A):i\in I\}$ be a
directed set and $(F,A)\leq \bigvee\limits_{i\in I} (G_i,A)$. We
write $\bigvee\limits_{i\in I} (G_i,A)=(G,A)$. $\forall e\in A$, we
have $G(e)=\bigcap\limits_{i\in I} G_i(e)\subseteq F(e)$. Then
$$U-F(e)\subseteq U-G(e)=\bigcup\limits_{i\in I} (U-G_i(e)).$$ For
any an $x\in U-F(e)$, there is an $i(x)\in I$ such that $x\in
U-G_{i(x)}(e)$. Now we get a finite set $\{(G_{i(x)},A): x\in
U-F(e)\}$, because $U-F(e)$ is finite. By the directness of
$\{(G_i,A):i\in I\}$, there exists an $i^{(e)}\in I$ such that
$(G_{i(x)},A)\leq (G_{i^{(e)}},A)$ for all $x\in U-F(e)$. Thus $x\in
U-G_{i(x)}(e)\subseteq U-G_{i^{(e)}}(e)$. We obtain $U-F(e)\subseteq
U-G_{i^{(e)}}(e)$, that is, $G_{i^{(e)}}(e)\subseteq F(e)$.

Since $A$ is a finite set, it implies that $\{(G_{i^{(e)}},A): e\in
A\}$ is also finite. By the directness of $\{(G_i,A):i\in I\}$
again, there exists a $j\in I$ such that $(G_{i^{(e)}},A)\leq
(G_j,A)$ for all $e\in A$. We have $G_{j}(e)\subseteq
G_{i^{(e)}}(e)\subseteq F(e)$ for all $e\in A$. This implies
$(F,A)\leq (G_j,A)$. Thus $(F,A)\ll (F,A)$ in ${\cal S}_{A}$.

(2) ``only if": For all $e\in A$, $U-F(e)$ can be represented as the
supremum of $\{B_i:i\in I\}$, where $\{B_i:i\in I\}$ is a directed
family of all the finite subsets of $U-F(e)$, i.e.,
$U-F(e)=\bigcup\limits_{i\in I} B_i$. We define a family of soft
sets $(H_i,A)$ as follows:
$$H_i(\varepsilon)=\left\{
          \begin {array}{ll}
          U-B_i,&{\mbox{if}} \ \ \varepsilon=e;\\
          F(\varepsilon),&{\mbox{otherwise}}.
         \end{array}
        \right.$$
With respect to the order relation $\leq$,
 $\{(H_i,A):i\in I\}$ is a directed subsets of ${\cal S}_{A}$.
Also we have $(F,A)= \bigvee\limits_{i\in I} (H_i,A)$ by Proposition
\ref{proposition:3}. Since $(F,A)\ll (F,A)$ in ${\cal S}_{A}$, there
exists a $k\in I$ such that $(F,A)\leq (H_k,A)$. Hence
$U-F(e)\subseteq U-H_k(e)=B_k$. Thus $U-F(e)$ is a finite subset of
$U$. This proves what we have stated. \qed

Let ${\cal S}_{\cal F}\subseteq {\cal S}$ denote the set of all soft
sets with a finite subset of $E$, i.e.,
$${\cal S}_{\cal F}=\{(F,A):
(F,A)\mbox{\ is\ a\ soft\ set\ over\ $U$}, \mbox{where\ $A$\ is\ a\
finite\ subset\ of\ $E$ } \}.$$ The symbol ${\cal P}_{f}(E)$ denotes
the set of all finite subsets of $E$.

Let ${\cal S}_{f,A}=\{(F, A): \forall e\in A, U-F(e)\mbox{\ is\ a\
finite\ subset\ of\ } U\}$. We denote
$${\cal S}_f=\mathop\bigcup\limits_{A\in {\cal P}_f(E)} {\cal S}_{f,A}.$$

\begin{theorem}\label{theorem:7}
{\rm $({\cal S}_{\cal F}, {\cal S}_f, {\cal P}_f(E))$ is a compact
information algebra. }
\end{theorem}

\noindent {\bf Proof.} First, we have $({\cal S}_{\cal F},{\cal
P}_f(E))$ is an information algebra. It is similar as the proof of
Theorem \ref{theorem:8}. By Proposition \ref{proposition:2}, we know
$({\cal S}_A,\leq)$ is a complete lattice for all finite subset $A$
of $E$. Hence the convergency in Definition \ref{definition:15} is
also true.

By the definition of way-below relation $\ll $ and the conclusion of
Lemma \ref{lemma:2}, the compactness is also clear.

Now we need to show the following equation holds for all finite
subset $A\subseteq E$,
$$(F,A)=\vee\{(G,A)\in {\cal S}_{f,A}: (G,A)\leq (F,A)\}.$$
For all $e\in A$, $U-F(e)$ can be represented as the supremum of
$\{B_i:i\in I^{(e)}\}$, i.e., $U-F(e)=\bigcup\limits_{i\in I^{(e)}}
B_i$, where $\{B_i:i\in I^{(e)}\}$ is a directed family of all the
finite subsets of $U-F(e)$. We define a family of soft sets
$(F_i,A)$ as follows:
$$F_i(\varepsilon)=\left\{
          \begin {array}{ll}
          U-B_i,&{\mbox{if}} \ \ \varepsilon=e;\\
          U,&{\mbox{otherwise}}.
         \end{array}
        \right.$$
Let ${\cal B}=\{(F_i,A):i\in I^{(e)},e\in A\}$. Clearly ${\cal
B}\subseteq {\cal S}_{f,A}$. Meanwhile, for all $(F_i,A)\in {\cal
B}$,  we have $(F_i,A)\leq (F,A)$. In fact, for all $e\in A$, if
$i\in I^{(e)}$, we have $F(e)=U-\bigcup\limits_{i\in I^{(e)}}
B_i\subseteq U-B_i=F_i(e)$. Otherwise, $F_i(e)=U$. Thus
$F(e)\subseteq F_i(e)$ is true. So $(F_i,A)\leq (F,A)$. We write
$\bigvee\limits_{(F_i,A)\in {\cal B}} (F_i,A)=(H,A)$. For all $d\in
A$, we have
\begin{eqnarray*}
 \begin{array}{lll}

H(d)&=&\bigcap\limits_{(F_i,A)\in {\cal B}} F_i(d)\\
&=&(\bigcap\limits_{i\in I^{(d)}} F_i(d))\cap (\bigcap\limits_{i\in I^{(e)},e\in A, e\not=d} F_i(d))\\
&=&(\bigcap\limits_{i\in I^{(d)}} F_i(d))\cap U\\
&=&\bigcap\limits_{i\in I^{(d)}} (U-B_i)\\
&=& U-\bigcup\limits_{i\in I^{(d)}}B_i\\
&=& F(d).
 \end{array}
 \end{eqnarray*}
So we have $F=H$. Therefore,
 \begin{eqnarray*}
 \begin{array}{lll}
 (F,A)&=&(H,A) \\
&=& \bigvee\limits_{(F_i,A)\in {\cal B}} (F_i,A)\\
 &\leq &\vee\{(G,A)\in {\cal S}_{f,A}: (G,A)\leq (F,A)\}\\
 &\leq & (F,A).
 \end{array}
 \end{eqnarray*}
Hence $$(F,A)=\vee\{(G,A)\in {\cal S}_{f,A}: (G,A)\leq (F,A)\}.$$

According to the proof above, we obtain that $({\cal S}_{\cal F},
{\cal S}_f, {\cal P}_f(E))$ is a compact information algebra. \qed

\section{Conclusion}\label{sec:con}
In this paper, by defining the operations combination and projection
of soft sets, we obtained the structure of information algebras on
the family of all soft sets over an initial universe set and a
parameter set. Therefore, a soft set system can be subsumed under
the specific instances of information algebra systems. We also gave
a model of compact information algebra in a soft set system. We have
shown the family of all soft sets with a finite parameter subset can
form a compact information algebra.


\section*{Acknowledgments}
This work is supported by National Science Foundation of China (Grant No.60873119) and the
Higher School Doctoral Subject Foundation of Ministry of Education of China under Grant 200807180005.

\end{document}